\documentclass[fleqn,usenatbib, letters]{mnras}

\usepackage{newtxtext,newtxmath}

\usepackage[T1]{fontenc}
\usepackage{ae,aecompl}

\newcommand{\code}[1]{\texttt{\detokenize{#1}}}

\usepackage{graphicx}	
\usepackage{amsmath}	
\usepackage{amssymb}	

\PassOptionsToPackage{pdfpagelabels=false}{hyperref}

\defcitealias{ElBadry_2018c}{ER18}

\title[ Wide binary metallicity dependence]{\vspace{-0.5cm} The wide binary fraction of solar-type stars: emergence of metallicity dependence at $a < 200$ AU \vspace{-0.5cm} }

\author[El-Badry \& Rix]{
Kareem El-Badry$^{1}$\thanks{E-mail: kelbadry@berkeley.edu}
and Hans-Walter Rix$^{2}$
\\
$^{1}$Department of Astronomy and Theoretical Astrophysics Center, University of California Berkeley, Berkeley, CA 94720\\
$^{2}$Max Planck Institute for Astronomy, D-69117 Heidelberg, Germany\\ \vspace{-0.5cm}
}

\date{Submitted to MNRAS Letters, September 18, 2018 \vspace{-0.5cm}}

\pubyear{2018}

\begin{document}
\label{firstpage}
\pagerange{\pageref{firstpage}--\pageref{lastpage}}
\maketitle

\begin{abstract}
We combine a catalog of wide binaries constructed from {\it Gaia} DR2 with [Fe/H] abundances from wide-field spectroscopic surveys to quantify how the binary fraction varies with metallicity over separations $50 \lesssim s/{\rm AU} \lesssim 50,000$. At a given distance, the completeness of the catalog is independent of metallicity, making it straightforward to constrain {\it intrinsic} variation with [Fe/H]. The wide binary fraction is basically constant with [Fe/H] at large separations ($s \gtrsim 250$\,AU) but becomes quite rapidly anti-correlated with [Fe/H] at smaller separations: for $50 < s/{\rm AU} < 100$, the binary fraction at $\rm [Fe/H] = -1$ exceeds that at $\rm [Fe/H] = 0.5$ by a factor of 3, an anti-correlation almost as strong as that found for close binaries with $a < 10$\,AU. Interpreted in terms of models where disk fragmentation is more efficient at low [Fe/H], our results suggest that $100 < a/{\rm AU} < 200$ is the separation below which a significant fraction of binaries formed via fragmentation of individual gravitationally unstable disks rather than through turbulent core fragmentation. We provide a public catalog of 8,407 binaries within 200 pc with spectroscopically-determined [Fe/H] for at least one component. 
\end{abstract}

\begin{keywords}
binaries: visual -- stars: formation -- stars: abundances 
\vspace{-0.2cm}
\end{keywords}



\section{Introduction}
The statistical properties of the binary star population and their potential variation with metallicity provide powerful diagnostics of the star formation process \citep{White_2001, Machida_2009, Duchene_2013, Bate_2014, Badenes_2018, Moe_2018}. Different binary formation mechanisms predict different metallicity dependences of the resulting binary fraction \citep{Moe_2018}. Measurements of metallicity dependence as a function of orbital separation can thus constrain the separation regimes in which different formation mechanisms operate.

The {\it close} binary fraction is strongly anti-correlated with metallicity: the fraction of primaries that have a companion with $a < 10$\,AU decreases by a factor of 4 over $-1 < \rm [Fe/H] < 0.5$, from $\left(40\pm 6\right)\%$ at $\rm [Fe/H] = -1$ to $\left(10\pm3\right)\%$ at $\rm [Fe/H] = 0.5$ \citep{Raghavan_2010, Badenes_2018, Moe_2018}. The existence of such anti-correlation was long controversial \citep{Jaschek_1959, Carney_1983, Latham_2002, Hettinger_2015}, in large part because the sensitivity of most binary detection methods varies with metallicity. But in a recent re-analysis of the binary populations probed by 5 surveys, \citet{Moe_2018} found that after observational biases are corrected for, all 5 show consistent evidence of a metallicity-dependent close binary fraction.

On the other hand, studies of the {\it wide} binary fraction have found it to be approximately metallicity-invariant \citep[e.g.][]{Zapatero_2004, Chaname_2004, ElBadry_2018b}. Most of the binaries studied in these works have separations of order 1000 AU, but the variation of metallicity dependence with orbital separation has yet to be studied in detail. The recent {\it Gaia} data releases \citep{Gaia_2016, Brown_2018} have substantially simplified the process of reliably identifying spatially resolved binaries, making it possible to study the intermediate-to-wide binary population with unprecedented precision \citep[e.g.][]{Oh_2017, Oelkers_2017, Andrews_2017}. In this Letter, we seek to constrain the transition between the close and wide binary regimes, pinpointing the separation at which metallicity dependence emerges. 

\section{Methods}
\label{sec:methods}

\subsection{Wide binary catalog and extension}
\label{sec:bin_cat}

We extend the wide binary catalog described in \citet[][hereafter ER18]{ElBadry_2018c}, which was constructed by searching {\it Gaia} DR2 for pairs of stars whose positions, proper motions, and parallaxes are consistent with being gravitationally bound. The \citetalias{ElBadry_2018c} catalog contains $\sim$50,000 binaries with separations $50 \lesssim s/{\rm AU} < 50,000$, with an estimated contamination rate of $\sim$0.2\%. To maintain high purity, it only contains binaries that are within 200\,pc of the Sun and havhe high-quality astrometry and photometry.

To separate main sequence stars and white dwarfs, \citetalias{ElBadry_2018c} required both stars to have a measured \code{bp_rp} color, and to have well-resolved photometry as quantified by the \code{phot_bp_rp_excess_factor} (see Section 2.1 of \citetalias{ElBadry_2018c}). As a consequence of these requirements, the \citetalias{ElBadry_2018c} catalog has an effective resolution limit of $\sim$2 arcsec and contains few binaries with separations $s< 200$\,AU. To find more binaries with small separations, we now extend the catalog by removing the restrictions on \code{bp_rp} and \code{phot_bp_rp_excess_factor} {\it only for pairs with projected separations $s < 500$}\,AU. We still require both components to pass the other quality cuts in \citetalias{ElBadry_2018c}, including having reliable astrometry and precisely measured, mutually consistent parallaxes, and we apply the same procedure for removing members of clusters, moving groups, and resolved higher-order multiples. 

This extended search yields 23,079 new binaries not included in the initial catalog. Many of the new additions have angular separations between 0.5 and 2 arcsec. Combining them with the sample from \citetalias{ElBadry_2018c} results in a total of 78,207 binaries, including 8,284 with $s<200$\,AU. Although the photometry of the objects that did not pass the \citetalias{ElBadry_2018c} cuts is less clean, we expect essentially all of them to be bona fide binaries, as the contamination rate from chance alignments is negligible at close separations. 

\subsection{Spectroscopic metallicities}
\label{sec:spec_met}
We cross-matched the expanded wide binary catalog with several wide-field spectroscopic surveys: LAMOST \citep[DR5;][]{Zhao_2012}, RAVE \citep[DR5;][]{Kunder_2017}, APOGEE (DR14; \citealt{Majewski_2017}, using the abundances derived by \citealt{Ting_2018}), and GALAH \citep[DR2;][]{Buder_2018}. We also cross-matched with the Hypatia catalog \citep{Hinkel_2014}, which is a compilation of high-resolution spectroscopic abundances for stars within 150 pc of the Sun. We limit our sample to main-sequence binaries in which the primary has absolute magnitude $2.5 < \rm M_G < 9.5$, corresponding to $0.45 \lesssim M/M_{\odot} \lesssim 1.5$.

\begin{table}
\caption{Sources of the spectroscopic abundances}
\label{tab:xmatch}
\begin{tabular}{lll}
\centering
Sample         & $N_{\rm in\,binary}$ & $N_{{\rm tot},\,d<200\,{\rm pc}}$ \\
\hline
RAVE           & 3,261     & 33,792  \\
LAMOST         & 3,729     & 46,729  \\
APOGEE         & 660       & 5,796   \\
GALAH          & 537       & 6,759   \\ 
Hypatia        & 660       & 3,954    \\

\hline
Total stars       & 8,847     & 97,030 \\
Total number of binaries & 8,407 & \\
\hline 
\end{tabular}
\end{table}

The resulting catalog is summarized in Table~\ref{tab:xmatch}. Cross-matching yields a spectroscopic [Fe/H] for at least one component of  8,407 binaries; in 440, a spectroscopic [Fe/H] is available for both components. We assign binaries in which both components have a spectroscopic [Fe/H] the mean of the two components; when only one component has a measured [Fe/H], we adopt that value. For stars that were observed by more than one survey, we prioritize abundances from surveys in the reverse order listed in Table~\ref{tab:xmatch}. The catalog is available online. 

We also construct a spectroscopic ``control sample'' that consists of the 97,030 stars within 200 pc that were observed by the spectroscopic surveys listed in Table~\ref{tab:xmatch} and pass the {\it Gaia} quality and magnitude cuts applied to the wide binary sample. Our method for identifying wide binaries is metallicity-blind, and the spectroscopic surveys did not preferentially target or avoid wide binaries. Therefore, any metallicity bias in the spectroscopic binary sample will, at fixed distance, affect the control sample in the same way it affects the binary sample. 

Binaries with angular separations of less than a few arcsec may not be spatially resolved by ground-based spectroscopic surveys. The resulting errors in [Fe/H] are expected to be less than 0.1 dex on average, with negligible systematic biases for the mid-resolution optical spectra that constitute the majority of our sample \citep{Schlesinger_2010, ElBadry_2018a}.
 
\section{Results}
\subsection{Metallicity Distribution}
\label{sec:res}
\begin{figure*}
\includegraphics[width=\textwidth]{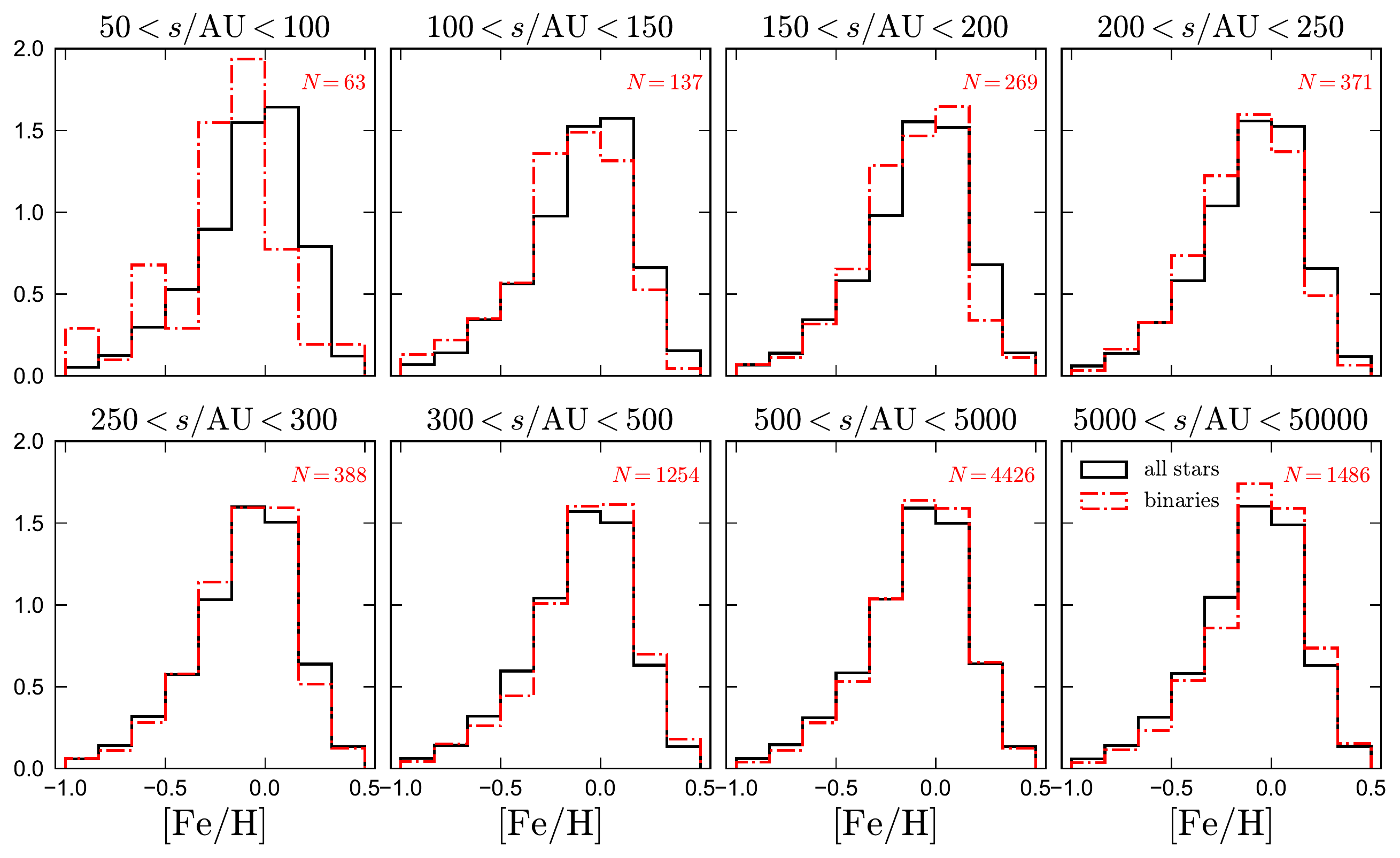}
\caption{Metallicity distribution functions (MDFs) of binaries (red) and all stars with spectroscopic metallicities (black). We separate binaries in bins of projected physical separation, $s$; in each panel, the black histogram shows the MDF of a random subset of the control sample with the same distance distribution as the binaries in that panel. At small separations ($s< 250$\,AU), the binary MDF shows a shortage of metal-rich binaries and an excess of metal-poor binaries relative to the control sample. At larger separations, the MDF of binaries is nearly identical to that of all stars, with a slight excess of metal-rich systems at the widest separations.} 
\label{fig:histograms}
\end{figure*}

Figure~\ref{fig:histograms} compares the metallicity distribution functions (MDFs) of binaries in different separation bins to the MDF of the control sample. Because close binaries are unresolved at large distances, their distributions of heliocentric distance are different from those of the full 200 pc control sample. To avoid biases arising from the distance-dependence of the MDF, for each bin in $s$, we compare to a random subset of the control sample with the same distance distribution as the binaries in that $s$ bin.

At small separations, the MDFs of binaries are biased toward low [Fe/H] relative to the control sample. This bias is strongest in the $50 < s/{\rm AU} < 100$ bin but is present in all separation bins up to $s = 250$\,AU. No strong bias toward higher or lower [Fe/H] is evident at large separations, though there are hints of a slight excess of binaries with $\rm [Fe/H]\sim 0$ in the largest separation bin. The latter is likely attributable to age effects: at the widest separations, there is a non-negligible probability for binaries to be dynamically disrupted by gravitational perturbations from other stars and molecular clouds. Lower-metallicity binaries are on average older, allowing more time for dynamical disruption.

\subsection{Inferring the dependence of binarity on [Fe/H]}
\label{sec:inference}
To quantify the metallicity dependence of the binary fraction implied by the binary MDFs in Figure~\ref{fig:histograms}, we fit a model that describes the ratio of the MDF of binaries to that of all stars as a function of metallicity and separation. Given a normalized, empirically measured MDF for all stars, $\psi\left({\rm \left[Fe/H\right]}\right)={\rm d}P/{\rm d}\left[{\rm Fe/H}\right]$, we define the MDF of binaries with separation $s$ as \begin{align}
\label{eq:psi_b}
\psi_{b}\left({\rm \left[Fe/H\right]}|s,\vec{m}\right)=\psi\left({\rm \left[Fe/H\right]}\right)w\left({\rm \left[Fe/H\right]}|s,\vec{m}\right)\times c.
\end{align}
Here $w$ is an arbitrary, positive-definite ``weight function'' parameterized by a vector of model parameters $\vec{m}$, and $c$ is a normalization constant, 
\begin{align}
\label{eq:c}
c\left(s,\vec{m}\right)=\left[\int\psi\left({\rm \left[Fe/H\right]}\right)w\left({\rm \left[Fe/H\right]}|s,\vec{m}\right)\,{\rm d\left[Fe/H\right]}\right]^{-1}.
\end{align}
If the MDF of all binaries is identical to that of single stars, then $w=\rm const.$ If the binary fraction varies with metallicity in way that affects binaries of all separations equally, $w$ is a function of $[\rm Fe/H]$ alone. In the general case where the binary fraction varies with [Fe/H] in a separation-dependent way, $w=w\left([{\rm Fe/H}], s\right)$. The absolute normalization of $w$ is arbitrary due to the normalization condition, but its dependence on [Fe/H] and $s$ determines the binary fraction relative to its value at a fixed metallicity; e.g. 
\begin{align}
\label{eq:fbin_ratio}
\frac{f_{{\rm bin}}\left(s,\left[{\rm Fe/H}\right]\right)}{f_{{\rm bin}}\left(s,\left[{\rm Fe/H}\right]=0\right)}=\frac{w\left(s,\left[{\rm Fe/H}\right]\right)}{w\left(s,\left[{\rm Fe/H}\right]=0\right)}.
\end{align}

We parameterize $w$ with a flexible function that allows the metallicity dependence to vary with separation:
\begin{align}
\label{eq:w_feh}
w=\frac{1}{2}\left[\left(w_0+1\right)-\left(w_0-1\right)\tanh\left(\frac{\left[{\rm Fe/H}\right]-\left[{\rm Fe/H}\right]_{0}}{\gamma}\right)\right].
\end{align}
This parameterization causes $w$ to asymptote to 1 at $[\rm Fe/H]\gg [\rm Fe/H]_0$ and asymptote to $w_0$ at $[\rm Fe/H]\ll [\rm Fe/H]_0$. The abruptness of the transition between the two regimes is determined by $\gamma$. To allow the asymptotic behavior to vary with $s$, we parameterize $w_0$ as 
\begin{align}
\label{eq:w0}
w_{0}=1+\frac{A-1}{1+\left(s/s_{0}\right)^{\beta}},
\end{align}
which asymptotes to $w_0=1$ at $s\gg s_0$ and to $w_0 = A$ at $s \ll s_0$. The abruptness of the transition is determined by $\beta$, which is required to be positive. 

Given a sample of binaries with metallicities $[{\rm Fe/H}]_i$ and separations $s_i$, the likelihood function is 
\begin{align}
L=p\left(\left\{ \left[{\rm Fe/H}\right]_{i},s_{i}\right\} |\vec{m}\right)=\prod_{i}\psi_{b}\left({\rm \left[Fe/H\right]}_{i}|s_{i},\vec{m}\right),
\end{align}
with $\psi_b$ given by Equation~\ref{eq:psi_b}. We tabulate the $\psi$ appropriate for the distance distribution of binaries as a function of $s$, using a subset of stars from the control sample with the same distance distribution as binaries of separation $s$. We sample the posterior distribution using \texttt{emcee} \citep{FormanMackey_2013}, with broad, flat priors on all model parameters $\vec{m}=\left(A,\log\left(s_{0}/{\rm AU}\right),\beta,\left[{\rm Fe/H}\right]_{0},\gamma\right)$.

\begin{figure*}
\includegraphics[width=\textwidth]{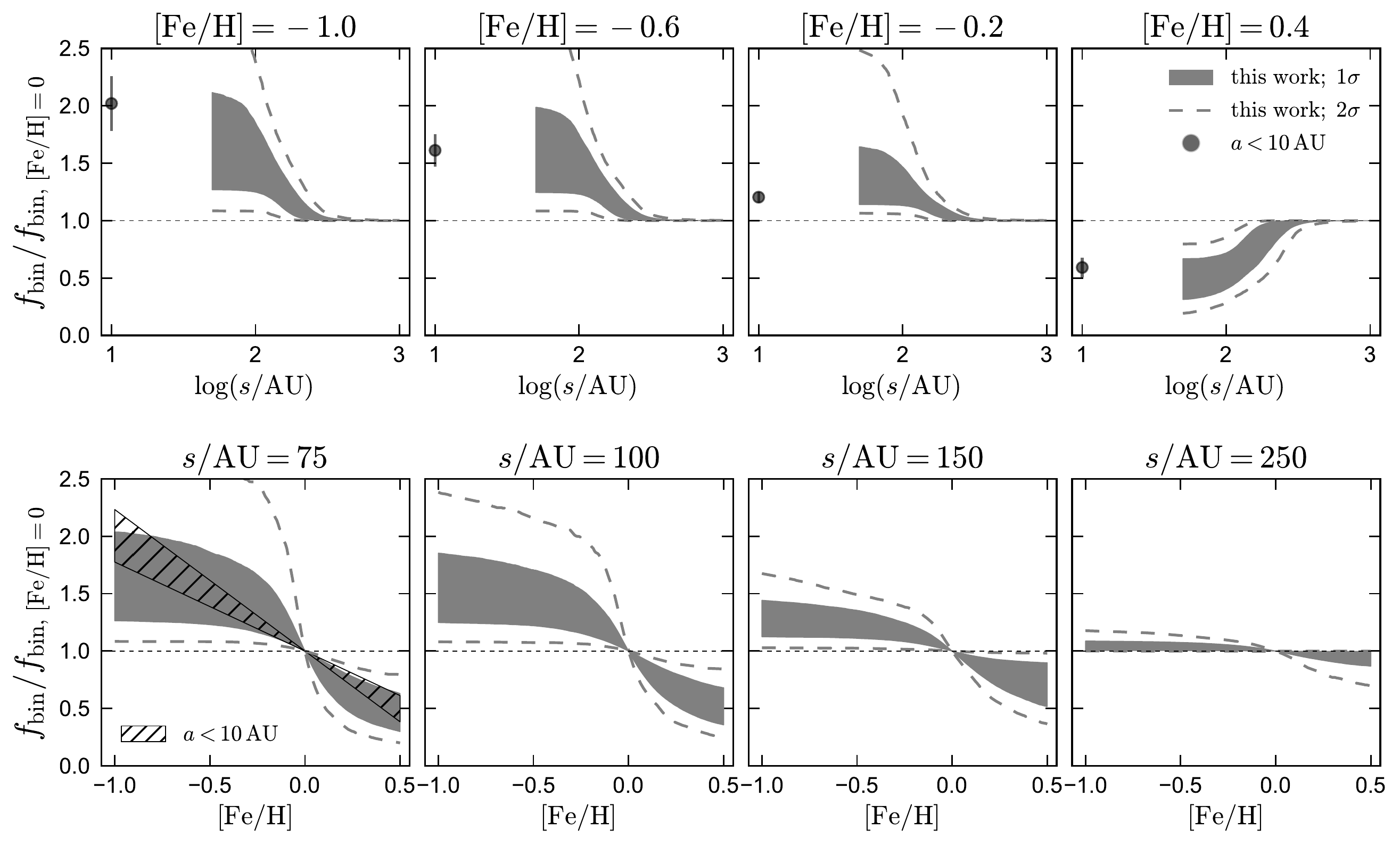}
\caption{Metallicity-dependence of the binary fraction. Top panels show constraints at fixed metallicity as a function of separation; bottom panels show constraints at fixed separation as a function of metallicity. In all panels, we normalize by the separation-dependent binary fraction at $[\rm Fe/H]=0$. Hints of metallicity-dependence first appear at $s\sim 250$\,AU and become stronger with decreasing separation. For comparison, we show constraints on the metallicity-dependence of the close binary fraction ($a < 10$\,AU) from \citet{Moe_2018}.}
\label{fig:projections}
\end{figure*}

We plot the resulting constraints on the binary fraction in Figure~\ref{fig:projections}, normalizing relative to $\rm [Fe/H]=0$. Consistent with the qualitative picture from Figure~\ref{fig:histograms}, the binary fraction becomes metallicity-dependent at $s\lesssim 250$\,AU. Over $-0.5 < [\rm Fe/H] < 0.5$, the dependence on metallicity is nearly linear, with hints of flattening at lower [Fe/H]. The emergence of metallicity dependence towards smaller separations is quite rapid: there is essentially no dependence at $s=300$\,AU, while at $s=100$\,AU, the binary fraction at $[\rm Fe/H]= -1$ is a factor of 3 higher than at $\rm [Fe/H] = 0.5$. The metallicity dependence at the smallest separations probed by our catalog is fully consistent with that found at $a <10$\,AU by \citet{Moe_2018}, though our median constraints lean toward somewhat weaker metallicity dependence at $\rm [Fe/H] < -0.5$. The uncertainties in our constraints are substantial at small separations due to the small number of binaries in our catalog with $s < 100$\,AU. Nevertheless, the data rule out a metallicity-independent binary fraction with greater than $2\sigma$ significance for all $s<150$\,AU.

\subsection{Separation distributions}
\label{sec:sep_dist}
\begin{figure}
\includegraphics[width=\columnwidth]{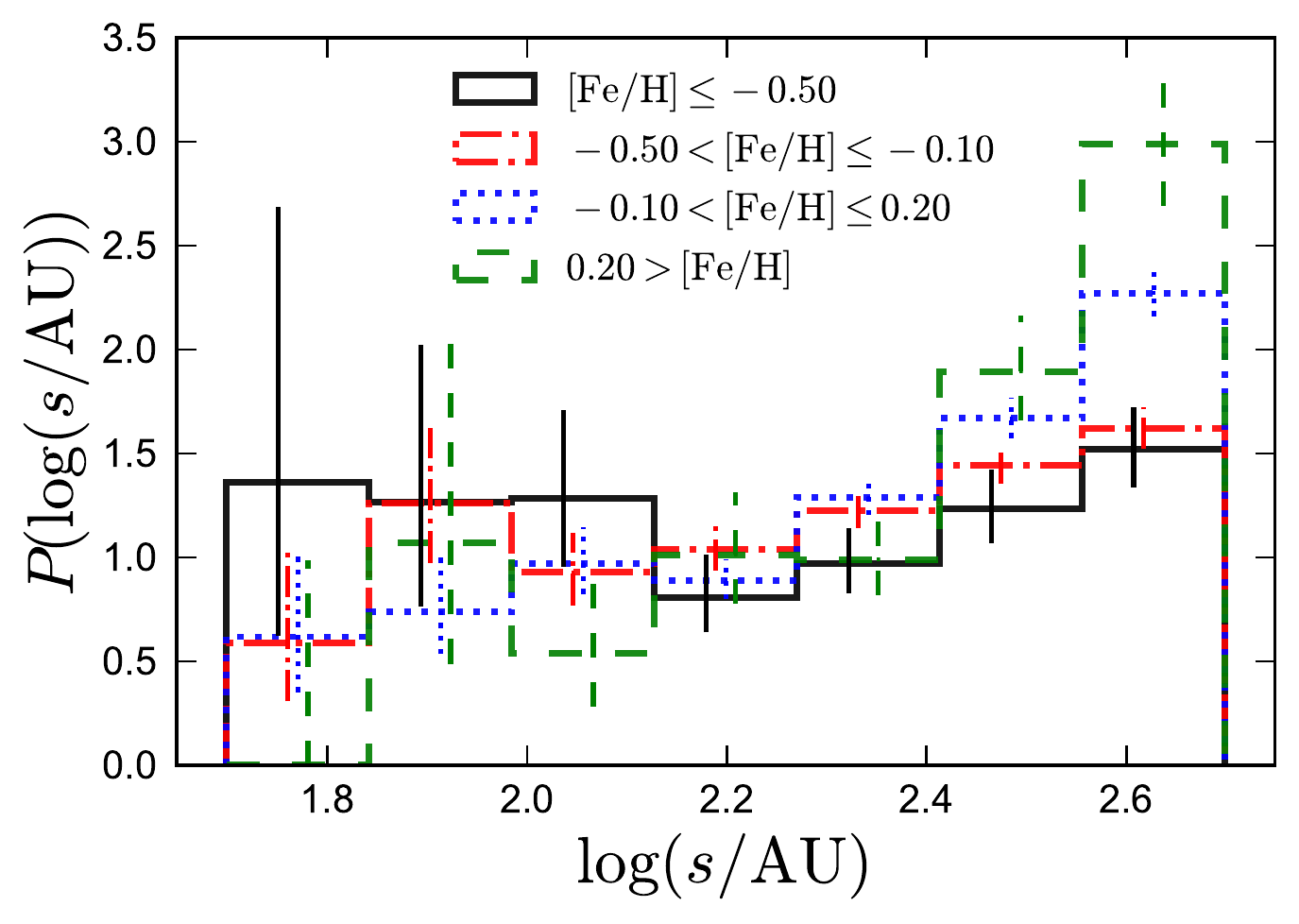}
\caption{Normalized separation distributions over $50<s/{\rm AU}<500$ for binaries of different metallicities, after correcting for incompleteness at small separations. Low-- (high--) [Fe/H] binaries are weighted toward smaller (larger) separations.}
\label{fig:sep_dist}
\end{figure}

In Figure~\ref{fig:sep_dist}, we show inferred intrinsic separation distributions over $50 < s/{\rm AU} < 500$ (where the binary selection function is consistent and well-characterized; see Section~\ref{sec:bin_cat}) of binaries in 4 metallicity bins. The relative detection efficiency drops below 50\% at an angular separation of $\theta_0 \approx 1$\,arcsec, so the maximum distance at which a binary of separation $s$ can be detected is $d_{{\rm max}}/{\rm pc}\approx{\rm min}\left\{ \left(s/{\rm AU}\right)/\left(\theta_{0}/{\rm arcsec}\right),\,200\right\}$. In inferring the intrinsic separation distribution, we weight each observed binary by the fraction of objects in the control sample at a particular [Fe/H] that are at distances greater than $d_{\rm max}$.

Over $50 < s/{\rm AU} < 500$, the separation distributions of lower-[Fe/H] systems consistently show an excess of smaller-separation binaries relative to those of higher-[Fe/H] systems. Integrating over all metallicities, the total separation distribution peaks at $a\approx 200$\,AU and is relatively flat over the separations shown in Figure~\ref{fig:sep_dist} \citep{Duchene_2013}. However, Figure~\ref{fig:sep_dist} shows that the peak in the separation distribution occurs at smaller (larger) separations for binaries with low (high) metallicity. Using a weighted two-sample KS test, we verified that the separation distributions for both of the two highest-[Fe/H] bins are inconsistent with those of each of the two lowest-[Fe/H] bins at at least the $4\sigma$ level ($p_{\rm KS}<3\times 10^{-5}$). Indeed, such variation is required to explain a binary fraction that is metallicity-dependent at small separations but not at asymptotically large separations (see \citealt{Moe_2018}, their Figure 19).

\section{Discussion and Conclusions}
\label{sec:discussion}

We have shown that the metallicity distribution functions (MDFs) of binaries with separations $50 \lesssim s/{\rm AU} \lesssim 250$ exhibit a shortage of high-[Fe/H] binaries and an excess of low-[Fe/H] binaries relative to a control sample subject to the same selection function (Figure~\ref{fig:histograms}). We fit a flexible, parameterized model for the modification of the binary MDF relative to the MDF of all stars, thus constraining the [Fe/H]-dependence of the binary fraction as a function of separation (Figure~\ref{fig:projections}). Metallicity dependence is weak at $s=250$\,AU but ramps up rapidly with decreasing separation: at $50 < s/{\rm AU} < 100$, the binary fraction increases by a factor of 3 over $-1 < \rm [Fe/H] < 0.5$. For these separations, the metallicity-dependence is roughly linear over $-0.5 < [\rm Fe/H] < 0.5$. It begins to flatten at $[\rm Fe/H] < -0.5$, albeit with substantial uncertainty. The separation distribution is similarly metallicity-dependent, with low-metallicity binaries concentrated at smaller separations (Figure~\ref{fig:sep_dist}).

The projected separation $s$ of a wide binary can exceed the semi-major axis $a$ by at most a factor of 2. Projection effects will tend to smooth out the transition between the metallicity-dependent and independent regimes, so the weak metallicity dependence detected at $s \gtrsim 200$\,AU could be a consequence of stronger metallicity dependence at $a\sim 150$. However, $a$ and $s$ are usually similar for realistic orbits (see \citetalias{ElBadry_2018c}; their Figure B1). For a uniform eccentricity distribution, 16\% of randomly observed orbits satisfy $s > 1.33a$, and 2.3\% satisfy $s > 1.75a$. Metallicity dependence out to $s\sim 250$\,AU thus implies dependence to at least $a = 150$\,AU. 

In the model proposed by \citet{Moe_2018}, the binary fraction is metallicity-dependent only at separations where binaries formed primarily via the fragmentation of gravitationally unstable disks: the turbulent core fragmentation process that produces wider binaries is independent of metallicity, at least for the range of $\rm [Fe/H]$ considered here. Interpreted in terms of this model, our results suggest that $100 < a/{\rm AU} < 200$ is the separation below which a substantial fraction of solar-type binaries were formed via disk fragmentation. 

In simulations, disk fragmentation typically occurs at separations of $a/{\rm AU}\sim 50-100$ \citep{Burkert_1997, Machida_2009, Stamatellos_2009, Kratter_2010}, in agreement with observations that find Class 0 binary protostars formed by disk fragmentation to have typical separations of $s\sim 75$\,AU \citep{Tobin_2013, Tobin_2016}. After fragmentation, the orbital separation can decrease due to viscous dissipation or three-body dynamics \citep[e.g.][]{Moe_2018b} or increase, e.g. following accretion of high-angular momentum gas. Our results imply that systems formed by disk fragmentation constitute a significant fraction of the binary population out to $a\sim 200$\,AU, with wider systems forming primarily through turbulent core fragmentation. Other lines of evidence also point toward a change in the binary formation mechanism at $a\sim 200$\,AU. The masses of the components of solar-type binaries are correlated and inconsistent with random pairings from the IMF out to separations of 200 AU \citep{Moe_2017}. Similar trends with separation are found for the correlation in accretion rates of binary protostars \citep{White_2001} and the mutual inclination of orbits in hierarchical triples \citep{Tokovinin_2017}.

It remains important to systematically measure the metallicity dependence of the binary fraction at intermediate separations ($10 \lesssim a/{\rm AU} \lesssim 50$), between the regimes probed by close binaries \citep[e.g.][]{Badenes_2018, Moe_2018} and our study. Some previous works have found metallicity-dependence in the binary fraction in this regime to be weak \citep{Raghavan_2010, Rastegaev_2010}, and others have found evidence of a {\it positive} correlation with metallicity, at least for low-mass binaries \citep{Riaz_2008, Jao_2009, Lodieu_2009, Ziegler_2015}. However, the common strategy of photometrically selecting metal-poor subdwarfs from below the main sequence for follow-up imaging can lead to a bias against binaries \citep{Moe_2018}. While speckle interferometry and AO + {\it HST} imaging likely represent the most promising route to probing the binary fraction at $10 \lesssim s/{\rm AU} \lesssim 50$, studies based on spectroscopically selected metal-poor samples would be less prone to biases against binaries. Constraints on the wobble of astrometric binaries in future {\it Gaia} data releases will also probe the intermediate-separation regime.

\section*{Acknowledgements}
We thank the anonymous referee for constructive comments, and Max Moe and Carles Badenes for helpful discussions.
KE was supported by the NSF GRFP. This project was developed in part at the 2018 NYC Gaia Sprint, hosted by the Center for Computational Astrophysics of the Flatiron Institute in NYC, and in part at the workshop ``Dynamics of the Milky Way System in the Era of Gaia,'' hosted at the Aspen Center for Physics, which is supported by NSF grant PHY-1607611. This work has made use of data from the ESA {\it Gaia} mission, processed by the {\it Gaia} Data Processing and Analysis Consortium (DPAC).  

The GALAH survey is based on observations made at the Australian Astronomical Observatory, under programmes A/2013B/13, A/2014A/25, A/2015A/19, A/2017A/18. We acknowledge the traditional owners of the land on which the AAT stands, the Gamilaraay people, and pay our respects to elders past and present. Funding for the Sloan Digital Sky Survey IV has been provided by the Alfred P. Sloan Foundation, the U.S. Department of Energy Office of Science, and the Participating Institutions. SDSS-IV acknowledges support and resources from the Center for High-Performance Computing at the University of Utah. The SDSS web site is www.sdss.org. Funding for RAVE (www.rave-survey.org) has been provided by institutions of the RAVE participants and by their national funding agencies. Guoshoujing Telescope (the Large Sky Area Multi-Object Fiber Spectroscopic Telescope LAMOST) is a National Major Scientific Project built by the Chinese Academy of Sciences. Funding for the project has been provided by the National Development and Reform Commission. LAMOST is operated and managed by the National Astronomical Observatories, Chinese Academy of Sciences. The research shown here acknowledges use of the Hypatia Catalog Database, an online compilation of stellar abundance data as described in Hinkel et al. (2014, AJ, 148, 54), which was supported by NASA's Nexus for Exoplanet System Science (NExSS) research coordination network and the Vanderbilt Initiative in Data-Intensive Astrophysics (VIDA).

\bibliographystyle{mnras}

\bsp	
\label{lastpage}
\end{document}